\documentclass{appolb}
\usepackage{epsfig}
\input epsf
\newcommand{\be} {\begin{equation}}
\newcommand{\ee} {\end{equation}}
\newcommand{\beqa} {\begin{eqnarray}}
\newcommand{\eeqa} {\end{eqnarray}}
\def\pmb#1{\setbox0=\hbox{#1}
\def\h{\hfill\break}
\def\half{\textstyle{1\over 2}}
\kern.05em\copy0\kern-\wd0 \kern-.025em\raise.0433em\box0 }
\def\pd{\partial}
\def\h{\hfill\break}

\font\sevenrm=cmr7
\begin{document}
\eqsec  
\title{Fundamental problems with \\ hadronic and leptonic interactions
\thanks{Presented at Epiphany meeting to commemorate Jan Kwiecinski, Krakow, January 2009}
}
\author{P V Landshoff
\address{University of Cambridge \\ pvl@damtp.cam.ac.uk}
}
\maketitle
\begin{abstract}
Common beliefs about unitarity are not reliable, and we do not know how to
apply DGLAP evolution at small x. Together with the big discrepancy between
the measurements of the total cross section at the Tevatron, 
a consequence is 
that the cross section at the LHC could be anywhere between 90 and 160~mb.
\end{abstract}
\PACS{11.55.Jy 13.60.Hb 13.85.Dz 13.85.Lg}
  
\section{Introduction}
The LHC will probe physics under extreme conditions. This means that past
physical intuition may be unreliable, and models that depend on intuition
may break down. Therefore it is important to ask which of our current beliefs
have a sound theoretical basis, and which are just based on folklore.

I will concentrate on two topics where our fundamental understanding
is particularly uncertain:\h
$\phantom{X}~~~~~\bullet ~$ unitarity\h
$\phantom{X}~~~~~\bullet ~$ DGLAP evolution at small $x$.\h
I will show that our lack of understanding of these has
serious consequences for what is probably the first thing that will be measured
at the LHC, the total cross section. The best estimate has a huge error:
\be
\sigma^{\hbox{{\sevenrm LHC}}}=125 \pm 35~\hbox{mb}
\label{sigtot}
\ee

My understanding of most of the material that I review here derives from my 
work over the years with Sandy Donnachie. Further details may be found in our 
book\cite{book}.
\section{Unitarity}
For an elastic hadron-scattering amplitude, the unitarity equation reads
\begin{center}
\epsfxsize=0.7\hsize\epsfbox{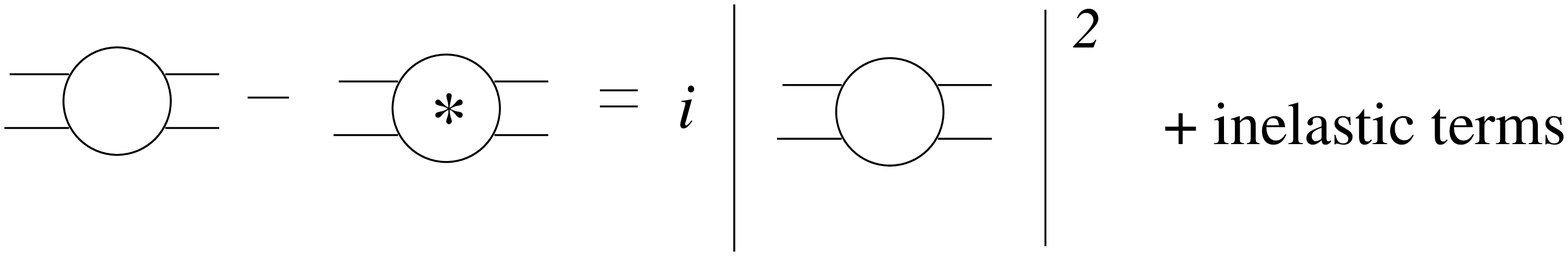}
\end{center}
or
\be
\hbox{Im }a_\ell (s)=|a_\ell (s)|^2 ~+~\hbox{inelastic terms}
\label{unitarity}
\ee
so that the partial-wave amplitude obeys
\be
|a_\ell (s)|<1
\label{pw}
\ee

A well-known consequence of (\ref{pw}) is the Froissart-Lukaszuk-Martin 
bound\cite{Fro61}\cite{LM67}:
\be
\sigma^{\hbox{{\sevenrm TOT}}}(s)<{\pi\over m_{\pi}^2}~\log^2(s/s_0)
\label{froissart}
\ee
At LHC energies, for reasonable values of the unkown scale $s_0$, this
gives a bound of several barns, and so it is not a useful constraint.

\def\P{I\!\!P }
A more useful bound is that of Pumplin\cite{Pum73}:
\be
\sigma^{\hbox{{\sevenrm ELASTIC}}}<\half ~\sigma^{\hbox{{\sevenrm TOTAL}}}
\label{pumplin}
\ee
The exchange of a single pomeron exchange $\P$ gives
$$
\sigma^{\hbox{{\sevenrm TOTAL}}}\sim s^{\epsilon}~~~~\epsilon\approx 0.08
$$
and
\be
{d \sigma\over dt}^{\hbox{{\sevenrm ELASTIC}}}\Big |_{t=0}
\sim s^{2\epsilon}
\label{singlepom}
\ee
which clearly causes this bound to be violated
at large enough $s$. The remedy is to
sum single-$\P$, double-$\P$, \dots\  exchanges:
\begin{center}
\epsfxsize 0.9\hsize\epsfbox{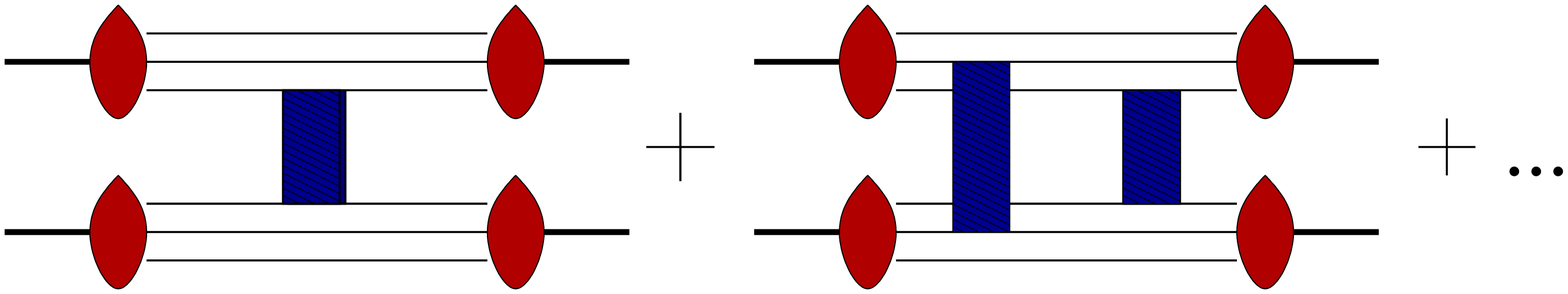}
\end{center}
However, even though we started trying to learn how to do this
more than 40 years ago, we still do not know how.

The best we can do, although it is certainly wrong\cite{book}, is to use an 
eikonal formalism. Write the amplitude as a 2-dimensional Fourier integral
\be
A(s, -{\bf q}^2)=4 \int d^2b\, e^{-i{\bf q}.{\bf b}} \tilde{A}(s,{\bf b}^2)
\label{eikonal}
\ee
Define $\chi(s,b)=-\log(1+2i\tilde A/s)$ so that
\be
\tilde{A}(s,{\bf b}^2)= \half is\big(1-e^{-\chi(s,b)}\big)
\label{chi}
\ee
Then the bound (\ref{pw}) on the partial-wave amplitude may be shown
to be equivalent to
\be
\hbox{Re }\chi(s,b)\ge 0
\label{constraint}
\ee
which is easy to impose.  If we expand the exponential in (\ref{chi})
in powers of $\chi$ we have 
\begin{eqnarray}
A(s, -{\bf q}^2)\!\! &=&\!\! 2is\int d^2b\, e^{-i{\bf q}.{\bf b}}
\big(1-e^{-\chi(s,b)}\big)\cr
\!\! &=&\!\! 2is\int d^2b\, e^{-i{\bf q}.{\bf b}} ~\Big(\chi-{{\chi^2}\over{2!}}
+{{\chi^3}\over{3!}}\dots\Big) 
\label{expand}
\end{eqnarray}

So far, the equations are certainly correct. But we do not know what to
take for $\chi(s,b)$. An obvious choice is to approximate it by 
single-$\P$ exchange. Although we do not know how to calculate
double-$\P$ exchange, we do know something about its general structure,
and the second term in (\ref{expand}) has the right structure. Similarly,
the third term has the right structure to represent triple-$\P$ exchange,
and so on. Nevertheless, this is wrong, for various reasons\cite{book}.
One one is that double-$\P$ exchange obviously depends on 
two-quark correlations in the proton wave function, which are absent 
in this procedure.

There has been a lot of talk about the consequences of unitarity 
for processes with leptons or photons in the initial state; for
example it is believed to lead to what is known as saturation. It is
important to understand that basic theory alone does not allow one to
conclude anything, without feeding in extra assumptions. That is, there
may well be no unitarity bound on $F_2(x,Q^2)$. The reason is that the 
analogue of (\ref{unitarity}) for $\gamma^*p$ scattering
\begin{center}
\epsfxsize=0.7\hsize\epsfbox{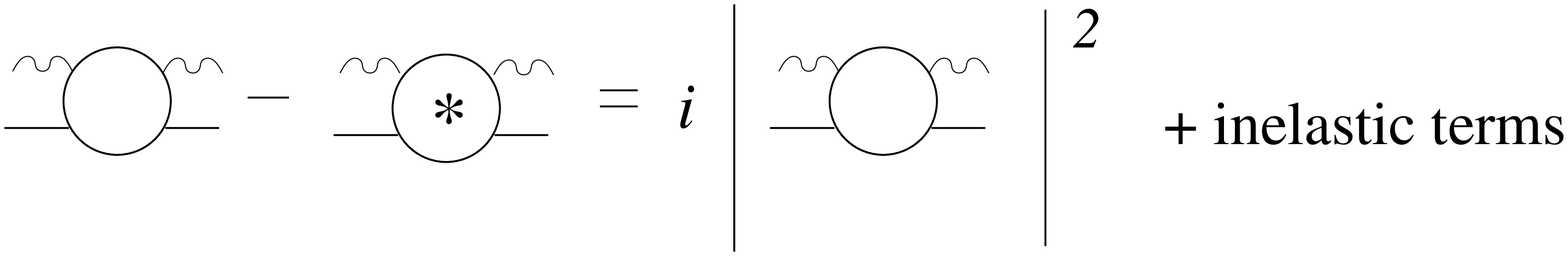}
\end{center}
is not true. This is the case even for $\gamma p$ scattering: for all we
know, the $\gamma p$ total cross section could continue to increase 
indefinitely with increasing energy.
\begin{figure}
\begin{center}
\epsfysize=0.6\vsize\epsfbox[65 295 377 755]{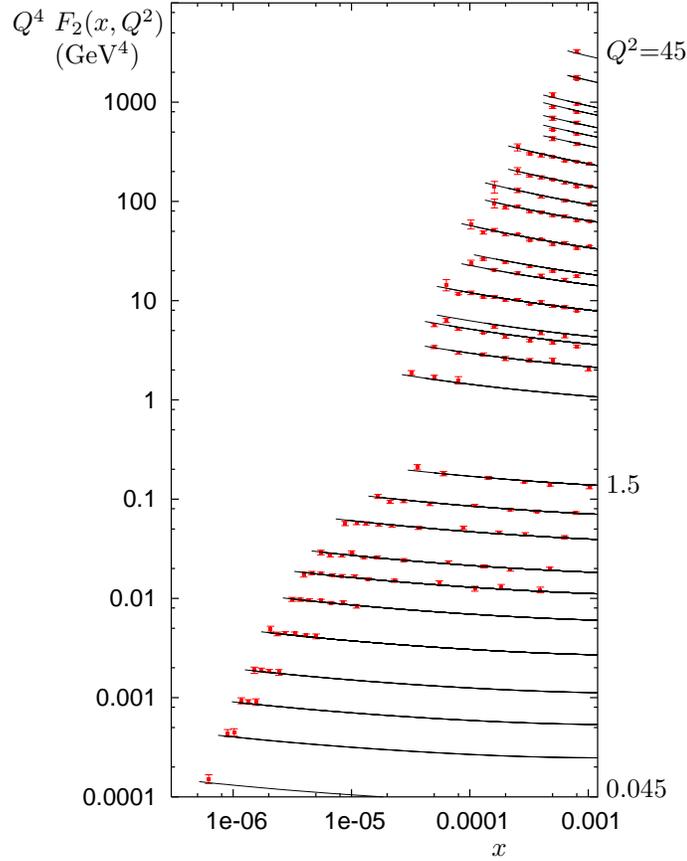}
\caption{Simple fit (\ref{simplest}) to data for $F_2(x,Q^2)$ at small $x$}
\label{simplest_fit}
\end{center}
\end{figure}

\section{DGLAP evolution at small $x$}
\def\u{{\bf u}}
\def\P{{\bf P}}
\def\f{{\bf f}}
Claims by experimentalists and theorists to be able to extract parton
densities with high accuracy from data, for use at the LHC, 
must be regarded with caution. This is because we do not know how
to handle DGLAP evolution at small $x$.

The singlet DGLAP eqation is:
\be
{\pd\over\pd t}\u (x,Q^2)=\int _x^1 dz\, \P (z,\alpha_s(Q^2))
\,\u({x/z},Q^2)
~~~~~~
\u(x,Q^2)=\left (\matrix{q(x,Q^2)\cr g(x,Q^2)\cr}\right ).
\label{dglap}
\ee
The terms of the perturbation expansion of  $\P (z,\alpha_s(Q^2))$ 
diverge like $1/z$ at $z=0$. For small $x$ the integration extends to
small values of $z$ and therefore it
is wrong to expand $\P (z,\alpha_s(Q^2))$ in powers of $\alpha_s$.

A possible exception is when $\u(x,Q^2)$ rises steeply with $1/x$.
To understand this, approximate $\u(x,Q^2)$ in some range of $x$ at
some value of $Q^2$ by
\be
\u(x,Q^2)\sim \f(Q^2)x^{-\epsilon}
\label{power}
\ee
and insert this into the DGLAP equation (\ref{dglap}). This gives
\be
{\pd\over\pd t}\log\f(Q^2)=\tilde\P(N=\epsilon, \alpha_s(Q^2))
-\int_0^x dz z^{\epsilon}\P (z,\alpha_s(Q^2))
\label{diffeq}
\ee
with $\tilde\P$ the Mellin transform of $\P$.
The last term $\sim x^{\epsilon}$ and so is negligible at small $x$
if $\epsilon$ is some way above 0.
A pole of $\P (z,\alpha_s(Q^2))$ at $z=0$ reflects itself in a pole
of $\tilde\P(N, \alpha_s(Q^2)$ at $N=0$. So if $\epsilon$ is some way above 0 
we are at some distance from this pole and then also it should be safe to 
expand the first term in powers of $\alpha_s$. By doing so, and dropping the 
last term in (\ref{diffeq}), we obtain a simple differential
equation for $\f(Q^2)$.

The simplest fit to {$F_2$} at small {$x$}
is a combination of two powers of $x$, hard-pomeron and soft-pomeron:
$$
F_2(x,Q^2)=f_0(Q^2)x^{-\epsilon_0}+f_1(Q^2)x^{-\epsilon_1}
$$
\be
\epsilon_0\approx 0.4~~~\epsilon_1=0.0808
\label{simplest}
\ee
The fit has just 5 free parameters, including $\epsilon_0$. 
See figure \ref{simplest_fit}.

\begin{figure}
\begin{center}
\epsfxsize=0.45\hsize\epsfbox[73 563 330 770]{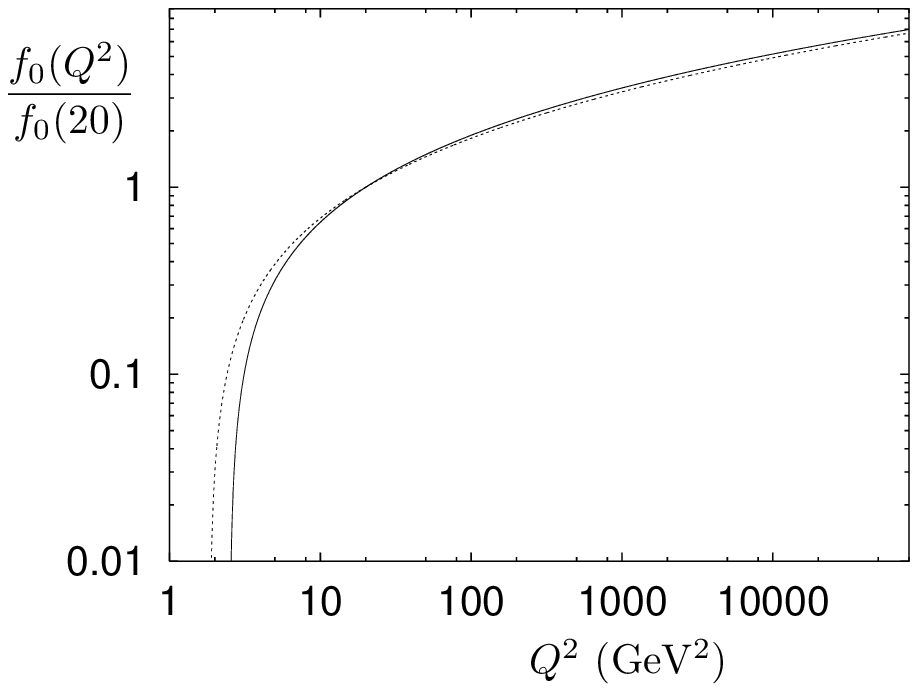}\hfill
\epsfxsize=0.45\hsize\epsfbox[73 563 330 770]{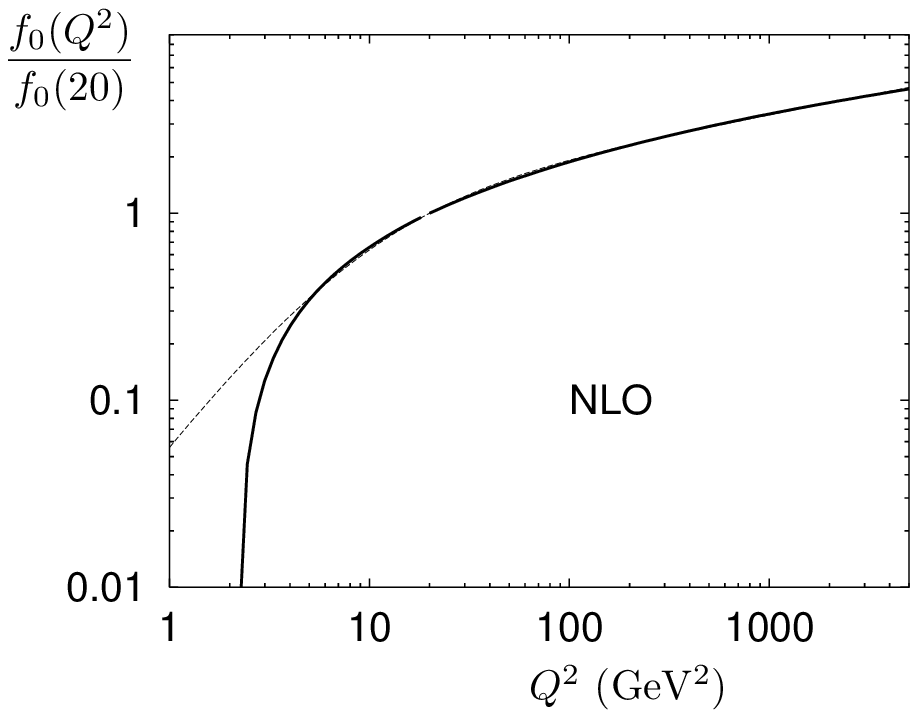}
\caption{Comparison of $f_0(Q^2)$ extracted from data with DGLAP evolution at LO and at NLO}
\label{evol}
\end{center}
\end{figure}

Donnachie and I made this fit\cite{DL98} purely phenomenogically, but
we then found\cite{DL02} that its output for the coefficient function
$f_0(Q^2)$ in (\ref{simplest}) obeys the DGLAP evolution differential
equation to very high accuracy, both at LO and at NLO. See figure
\ref{evol}. 
DGLAP evolution is supposed to be valid only for large
$Q^2$ and we can see from the figure that this means at least 5 to 10
GeV$^2$. It certainly does not make sense to use it down to 1 or 2
GeV$^2$, as is often done.

As I have explained, DGLAP evolution breaks down for
the soft-pomeron coefficient function $f_1(Q^2)$, as $\epsilon_1$ is
too close to~0. The conventional approach to DGLAP evolution, used by
many theorists and experimentalists to extract what are claimed to be
highly accurate parton distributions, amounts to ignoring this
difficulty.

\begin{figure}
\begin{center}
\epsfxsize=0.45\hsize\epsfbox[125 495 470 770]{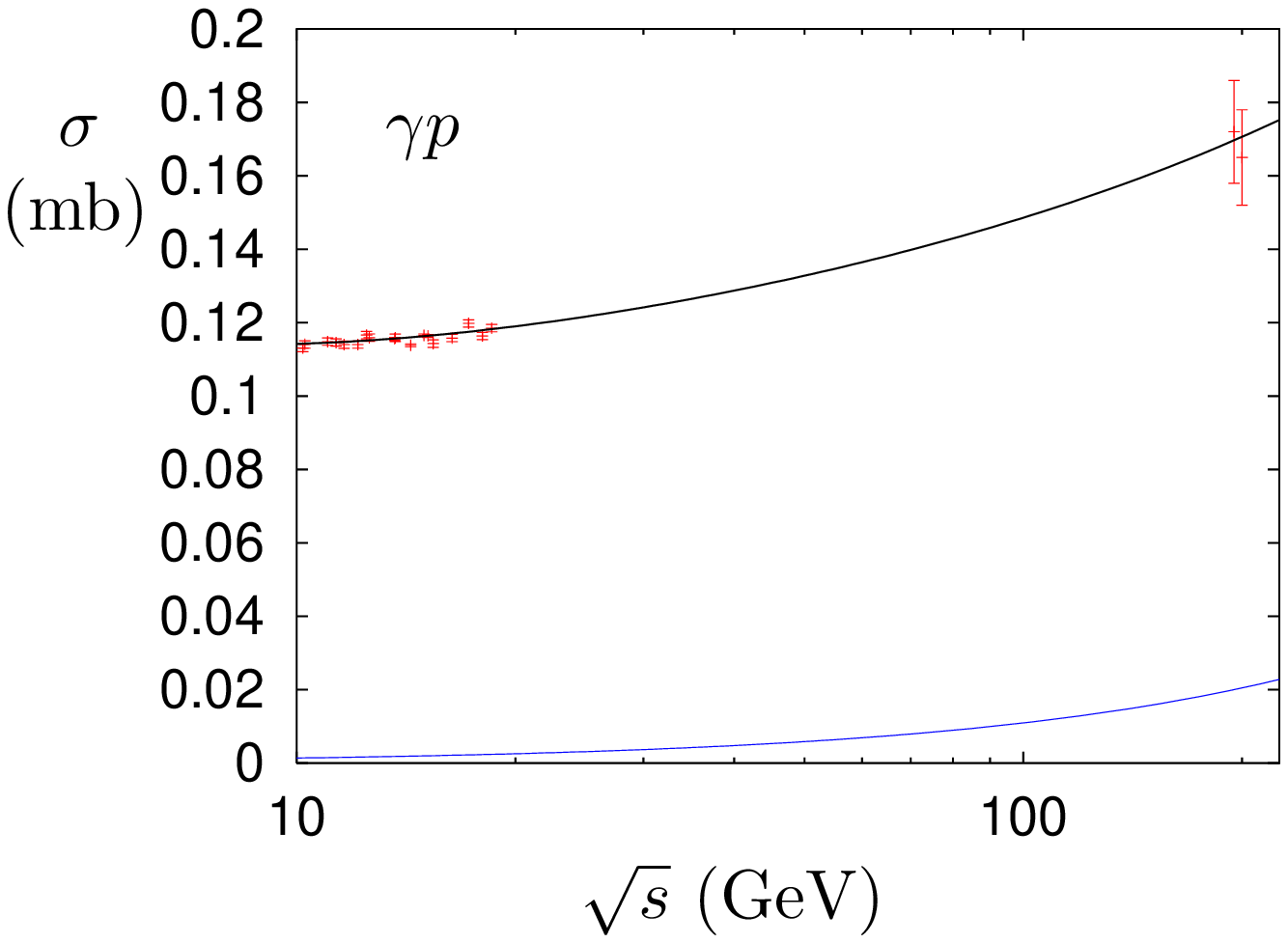}\ 
\epsfxsize 0.71\hsize\epsfbox[100 300 450 770]{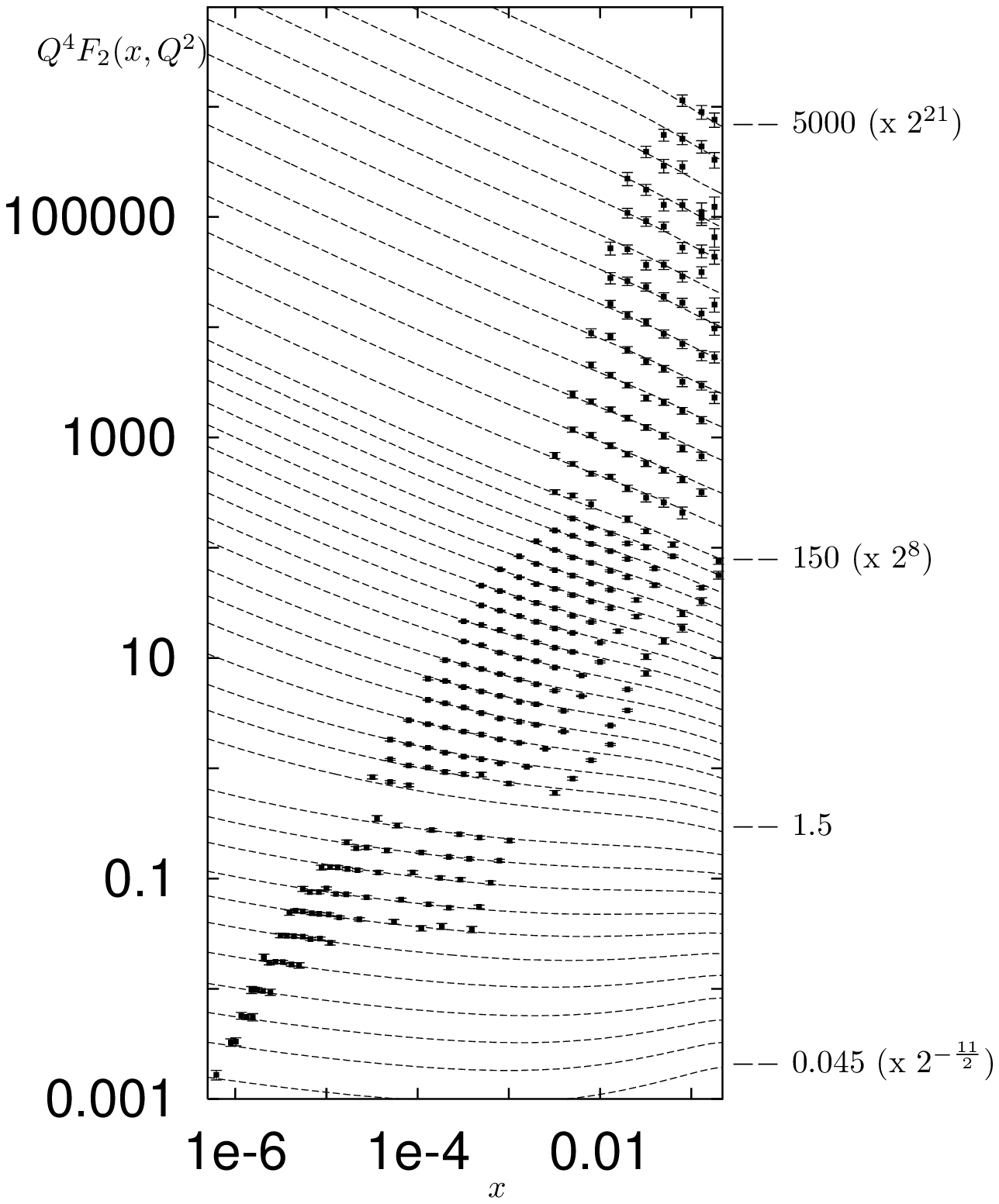}
\caption{The fit to $F_2$ extended to larger values of $x$, and to real-photon
data}
\label{astonish}
\end{center}
\end{figure}

Our simple fit (\ref{simplest}) applies only at small $x$, which is why
figure \ref{simplest_fit} shows data only for
$x<10^{-3}$.
If we want to extend it to larger values of $x$, we should 
add in a Regge term $x^{-\epsilon_R}$ with $\epsilon_R\approx -\textstyle{\half}$,
corresponding to $f_2$ and $a_2$ exchange. Also, the simple powers of
$x$ in (\ref{simplest})
must be multiplied by functions that go to 0 as $x\to 1$. We do
not know what these should be, so Donnachie and I took\cite{DL98}
just the powers of $(1-x)$ given by the dimensional counting rules. This is 
certainly too simple to be correct but is better than doing nothing. It is also
astonishingly successful: see figure \ref{astonish}, which shows
also what the fit gives for the $\gamma p$ total cross section.
Compared with figure \ref{simplest}, only 2 free parameters have been
added, both for the Regge term.

\section{Total cross section at the LHC}

\begin{figure}
\begin{center}
\epsfxsize=0.45\hsize\epsfbox[125 495 470 770]{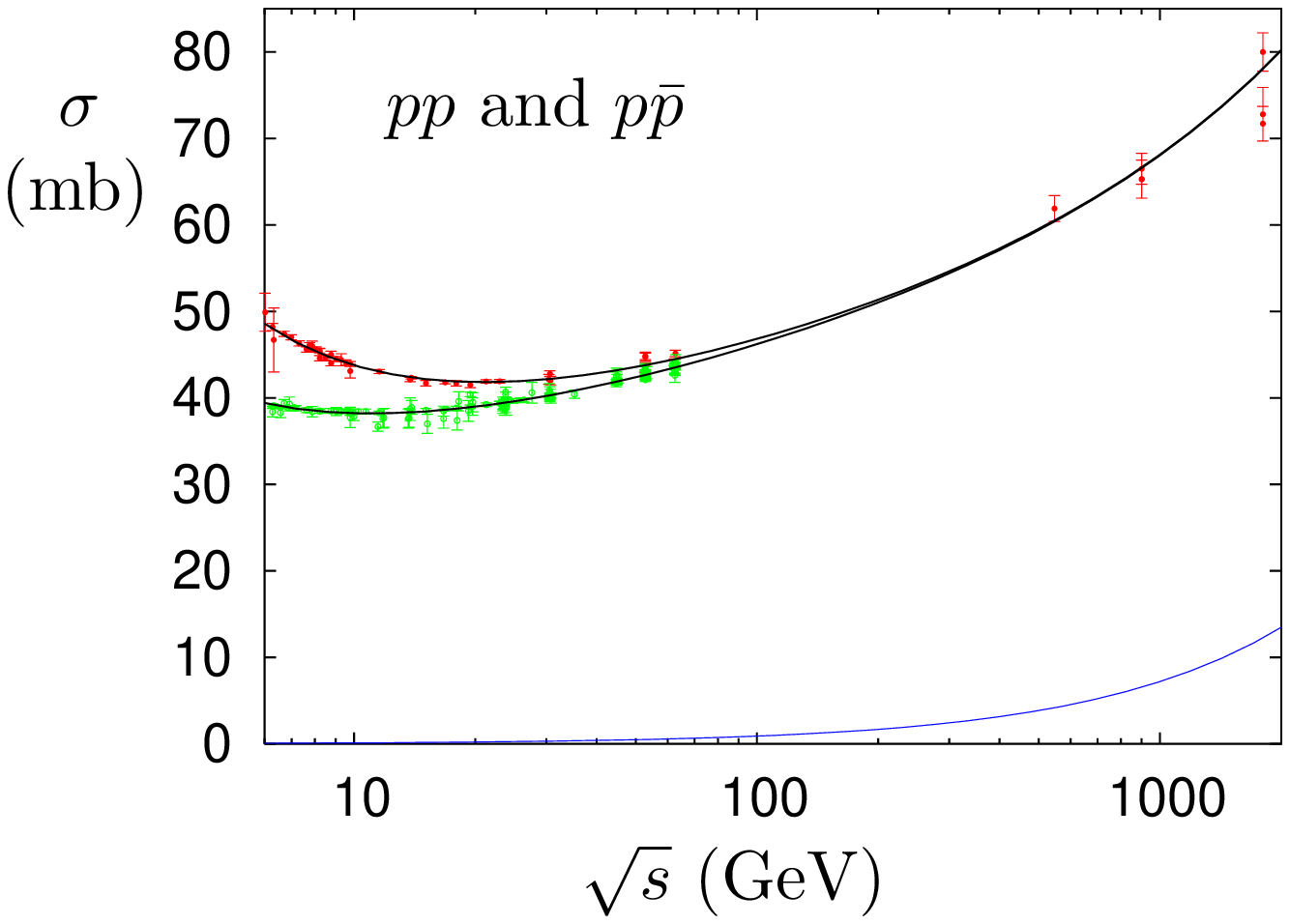}\hfill \epsfxsize=0.45\hsize\epsfbox[125 495 470 770]{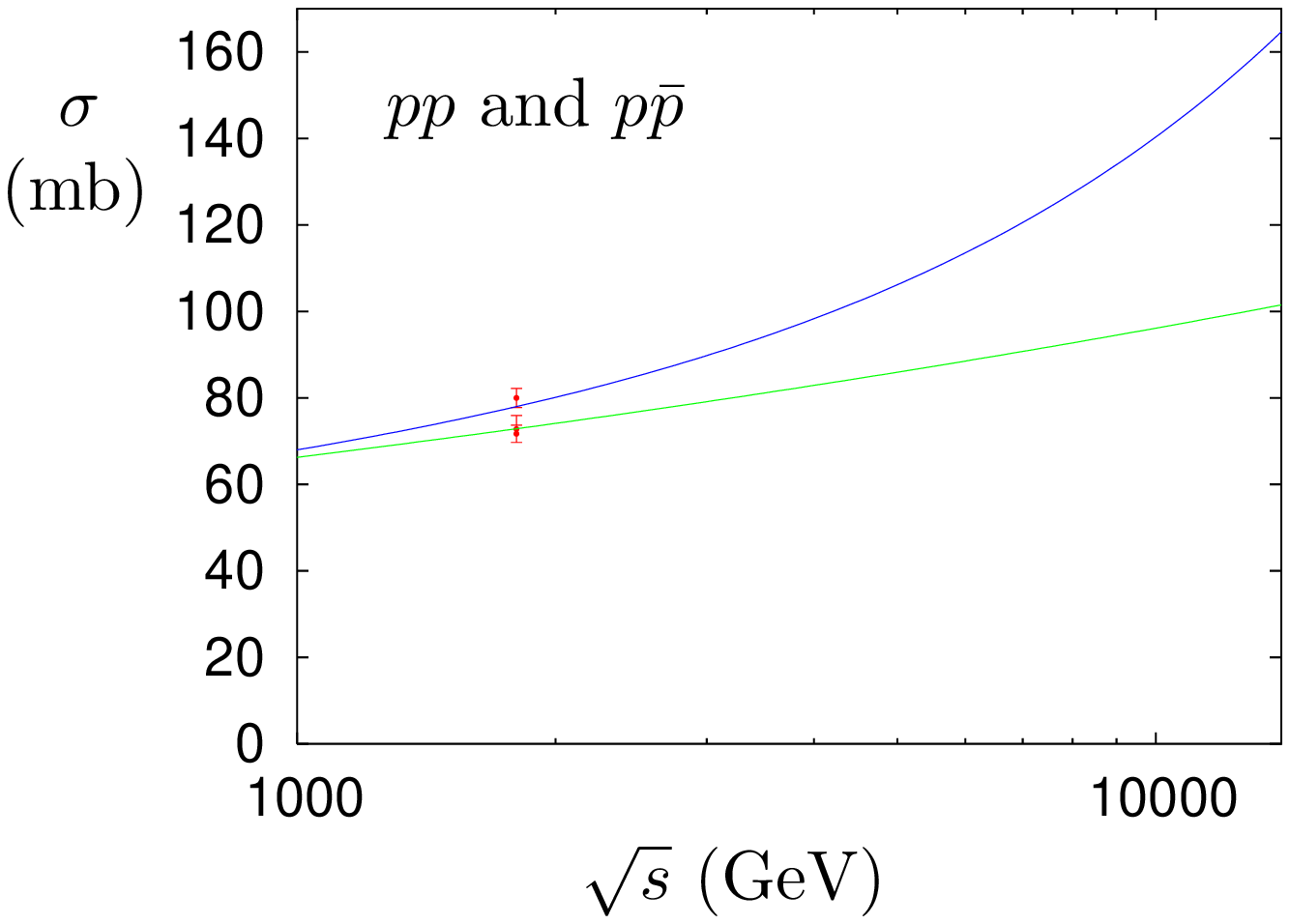}
\caption{Fits to $pp$ and $\bar pp$ total cross sections}
\label{factor}
\end{center}
\end{figure}

Given that the data for $F_2(x,Q^2)$ respond so well to a fit that includes
a hard pomeron, it is natural\cite{DL04} to include such a term also in fits to $pp$ and 
$\bar pp$ scattering. That is, for each of
$\sigma(pp),\sigma(p\bar p),\sigma(\gamma p)$ include hard pomeron,
soft pomeron and Regge exchange:
\be
\sigma=X_0s^{\epsilon_0}+X_1s^{\epsilon_1}+X_Rs^{\epsilon_R}
\label{hard}
\ee
The result is shown in figure \ref{factor}. In the left-hand figure, the
lowest curve is the hard-pomeron contribution. The right-hand figure shows the
extrapolation to LHC energy of the two fits, this new one and the old fit without a hard pomeron
term.

Notice the familiar and long-standing discrepancy between the E710 and CDF
measurements at the Tevatron. If the upper CDF measurement should be correct,
it surely is a sign that something new is beginning to become important
at Tevatron energy, with the consequence that the LHC cross section will
be large. This observation does not depend on any particular theoretical
explanation of the data. 

However, the large prediction for the LHC total cross section when the 
hard pomeron is included will lead some to worry about unitarity. 
Some even worry about unitarity for the lower curve.
Donnachie and I stressed when we made the original DL fit\cite{DL92} that
the power $\epsilon_1\approx 0.08$ was an effective power that already,
to some extent, includes unitarity corrections; nevertheless Alan
Martin and collaborators believe\cite{martin} that this is not enough
to take account of unitarity and so predict that the LHC cross section
will be about 90~mb.

I have explained that nobody knows how to calculate the effects of
unitarity. I will now describe an attempt to do so, which should
not be taken at all seriously.
\begin{figure}
\begin{center}
\epsfxsize=0.55\hsize\epsfbox[30 348 330 825]{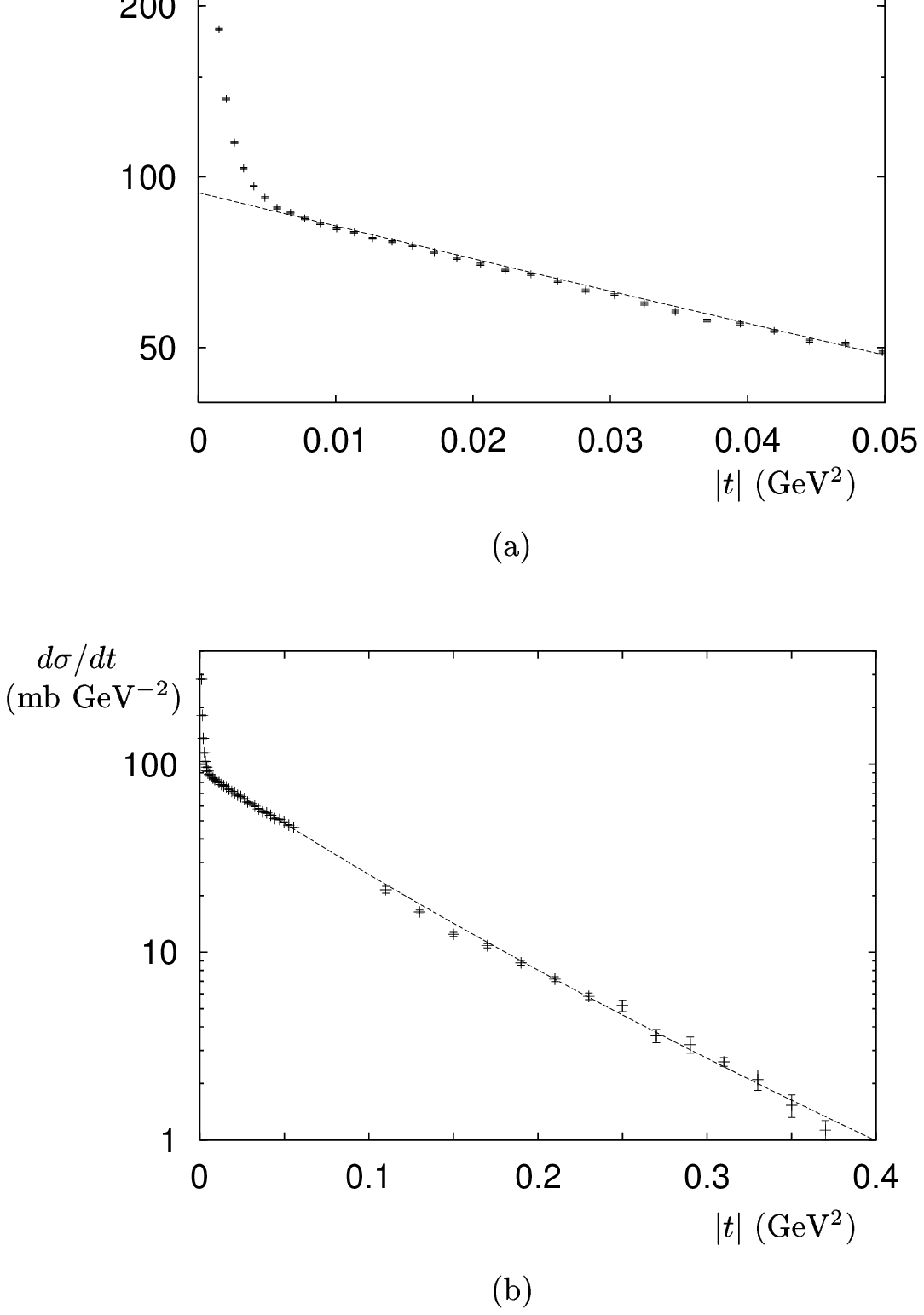}
\caption{$pp$ elastic scattering data at $\sqrt s=53$ GeV. The curves do not include photon exchange}
\label{el}
\end{center}
\end{figure}
\begin{figure}
\begin{center}
\epsfxsize=0.58\hsize\epsfbox[88 460 475 760]{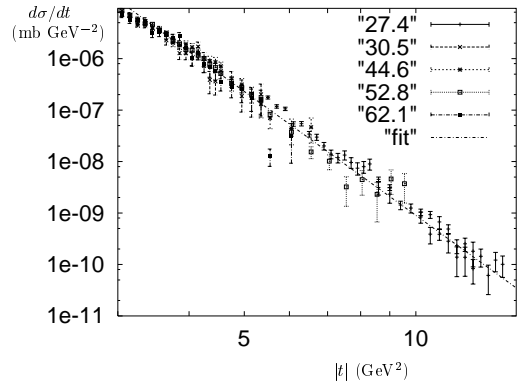}
\caption{Large-$t$ data for $pp$ elastic scattering at various energies,
with the fit (\ref{larget})}
\label{large_t}
\end{center}
\end{figure}
\begin{figure}
\begin{center}
\epsfxsize=0.2\hsize\epsfbox{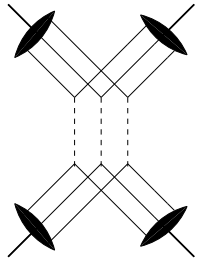}
\caption{Triple-gluon exchange}
\label{ggg}
\end{center}
\end{figure}

Consider $pp$ and $\bar pp$ elastic scattering. At existing energies,
soft pomeron exchange dominates and its contribution has been known
for nearly 40 years:
\be
{d\sigma\over dt}={[3 \beta _1F_1(t)]^4\over{4\pi}}  (\alpha' _1s)^{2(\epsilon _1 +\alpha' _1t)}
\label{elastic}
\ee
Here $F_1(t)$ is the elastic form factor of the proton and
$\beta _{1}$ and $\epsilon _{1}$ are known from 
$\sigma ^{\hbox{{\sevenrm TOT}}}$. As long ago as 1973, my then
student Jaroskiewicz\cite{Jar74} found 
that data fix the only free parameter $\alpha' _{1}$ to be
0.25 GeV$^{-2}$. This is done by using very-small-$t$ data at some energy: see
the upper part of figure (\ref{el}). Then the formula (\ref{elastic}) fits the
data out to rather larger values of $t$ at that energy, as is seen in in
the lower plot in the figure. Because $F_1(t)$ is raised to the 4th power
the fit is rather sensitive to it. I do not understand why the proton's 
electromagnetic form factor should be appropriate, since pomeron exchange has 
the opposite $C$ parity from photon exchange.

As is well known, at small $t$ $pp$ elastic scattering displays shrinkage:
$d\sigma /dt$ becomes steeper as the energy increases. The rate of
shrinkage is determined by the value of $\alpha' _{1}$, and 0.25 GeV$^{-2}$
describes the data very well at all the different energies that have 
been measured\cite{book}.

At large values of $|t|$, greater than about 3 GeV$^2$, the data take on
a different character. They are described very well by
\be
{d\sigma\over dt} = 0.09\,t^{-8}
\label{larget}
\ee
and are independent of energy. See figure \ref{large_t}. The form 
(\ref{larget}) is what is obtained from the triple-gluon-exchange
mechanism of figure \ref{ggg}. This raises an interesting question\cite{DL96}:
what if one replaces each gluon with a hard pomeron? This might provide
a mechanism which, while it is too small to be seen at existing energies,
grows rapidly with energy and is large at LHC energy. That is, $d\sigma/dt$
at the LHC might be large at large $t$.

\begin{figure}
\begin{center}
\epsfxsize=.49\hsize\epsfbox[35 575 320 775]{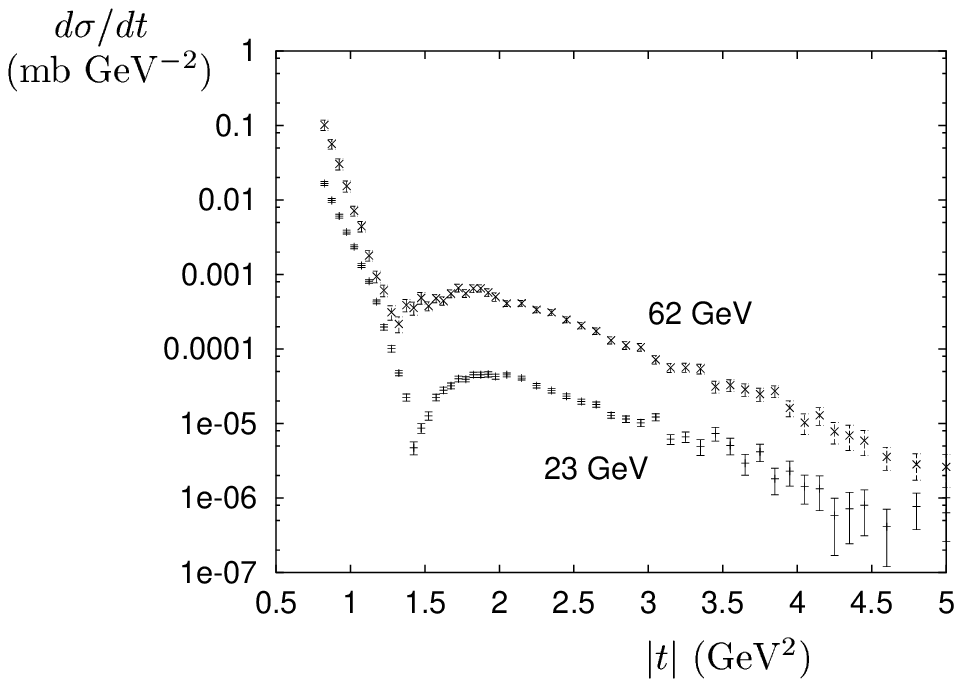}\hfill
\epsfxsize=.45\hsize\epsfbox[50 580 310 770]{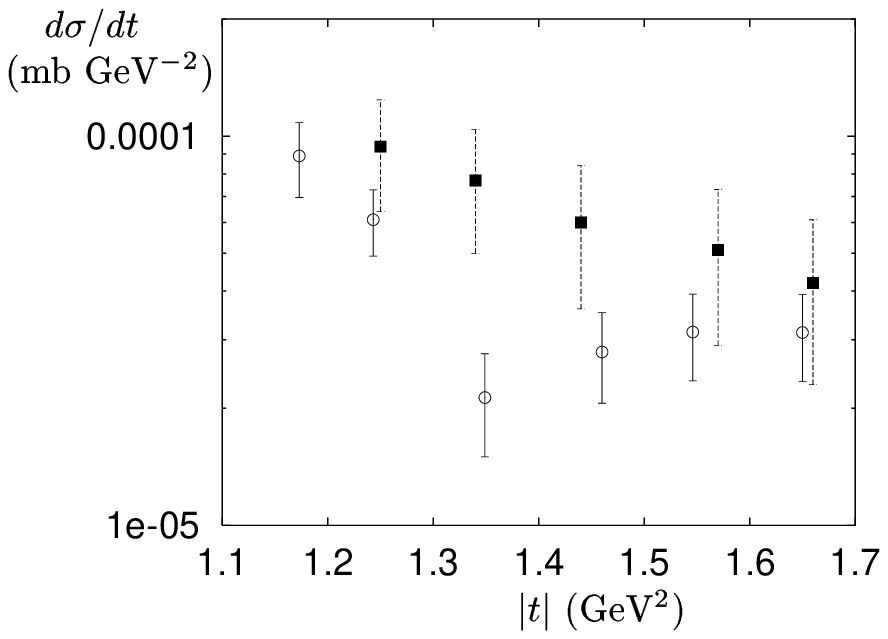}
\caption{$pp$ and $\bar pp$ elastic scattering data. The left hand figure 
is for $pp$ and the 62 GeV data are multiplied by 10. The data in the 
right-hand figure is at 53 GeV and the upper points are $\bar pp$, the 
lower $pp$.}
\label{dip}
\end{center}
\end{figure}
\def\P{I\!\!P }
At energies corresponding to the data plotted in figures \ref{el} and 
\ref{large_t}, $pp$ elastic scattering data display a striking dip structure 
at values of $t$ in between those of the two figures. See figure \ref{dip}.  
It is not easy to generate a dip: the real and imaginary parts of the
amplitude must be very small at the same $t$, which requires something
of an accident. It needs at least three
contributions: probably $\P$, $\P\P$ and $ggg$. However, the third of these
three terms has the opposite $C$ parity from the first two, which
led us to predict\cite{DL83} that $\bar pp$ scattering should not
have a dip. As the figure shows, this was later confirmed. An exchange
such as $ggg$ which has negative $C$ parity is known as odderon exchange;
it is a mystery why odderon exchange has not been detected at smaller values of
$t$.
\begin{figure}
\begin{center}
\epsfxsize=0.5\hsize\epsfbox[50 50 410 300]{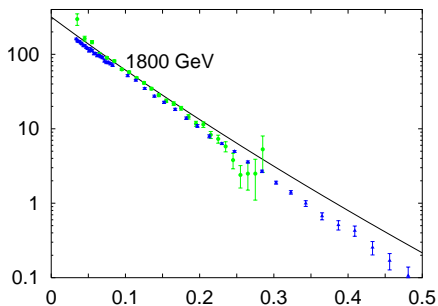}
\caption{$\bar pp$ elastic scattering data from the Tevatron; the curve corresponds to the exchange of single hard and soft pomerons}
\label{single_el}
\end{center}
\end{figure}

\begin{figure}
\begin{center}
\epsfxsize=0.95\hsize\epsfbox[80 560 530 770]{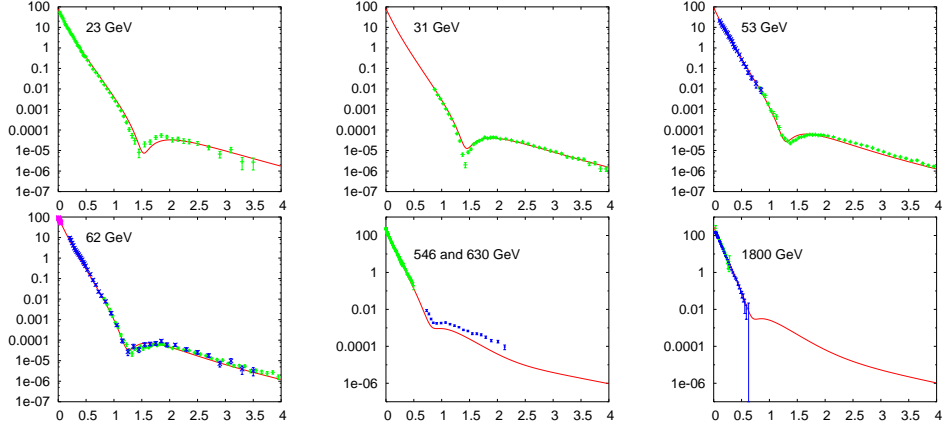}
\caption{$pp$ and $\bar pp$ elastic scattering data at various
with fit $\P +\P\P +ggg$}
\label{dsig}
\end{center}
\end{figure}

The fit shown in figure \ref{el} to elastic scattering at relatively low
energy remains good at very small $t$ as the energy is increased
but becomes less good at larger values of $t$.
Figure \ref{single_el} shows data for $\bar pp$ elastic scattering
from the two Tevatron experiments. The curve corresonds to the exchange of a 
single soft pomeron plus that of a single hard pomeron. I have explained that
we do not know how to calculate the exchange of two pomerons, but we 
know enough about its general features to know that including this will
pull  $d\sigma/dt$ down at larger $t$. As a very crude model that includes
the two-pomeron exchanges $\P_0\P_0, \P_1\P_1$ and $\P_0\P_1$ (where
again the subscripts 0 and 1 denote the hard and soft pomerons), I 
calculated the $b$-space amplitude for the sum of the $\P_0$ and
$\P_1$ exchanges,
and  squared it to simulate the sum of the $\P\P$ exchanges:
\be
\tilde A(s,b)=2is\Big (\chi(s,b)-\lambda [\chi(s,b)]^2\Big )
\label{twopom}
\ee
I chose the value of $\lambda$ to cancel the imaginary part of the amplitude so
as to get the dips in $pp$ scattering at
the right value of $t$, and added in $ggg$ exchange. The large-$t$ form
(\ref{large_t}) of the latter needs to be modified at smaller $t$
so that it does not diverge, and one does not know how to do this, but 
I guessed a form with a single parameter which I chose so as the optimise
the fit to the dip structure. My best
fit to the data is shown in figure \ref{dsig}.
\begin{figure}
\begin{center}
\epsfxsize=0.6\hsize\epsfbox[90 590 320 760]{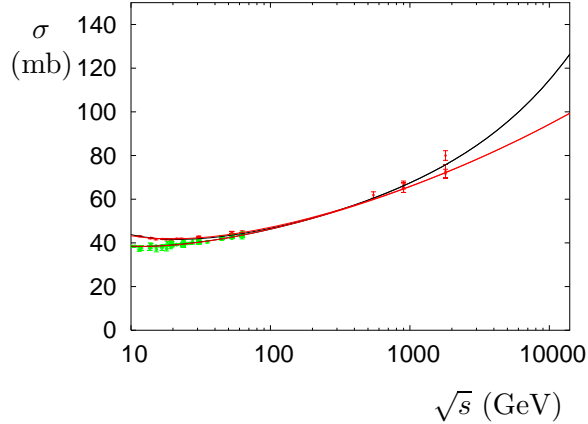}
\caption{Extrapolations to LHC energy of fits to the total cross section. 
The upper curve corresponds to the fits to the amplitude shown in figure
\ref{dsig}, including $\P\P$ exchanges, while the lower curve omits
any hard-pomeron contribution}
\label{pphardsoft}
\end{center}
\end{figure}

The result for the total cross section at the LHC is
that including the two-pomeron exchanges pulls the prediction
down from 160 to 125 mb: see figure \ref{pphardsoft}. The lower curve
in the figure corresponds to single soft pomeron exchange only. So, since
my attempt to include the double exchanges is surely very crude, I have to conclude that the LHC total cross section could be anywhere between 100 and 
160~mb. Remember, though, that it might be even smaller\cite{martin}.

\section{Summary}
\def\b{$\bullet~$}

\b $\sigma^{\hbox{{\sevenrm LHC}}}=125 \pm$ 35 mb

\b We do not know how usefully to impose unitarity --- eikonal-type models
are surely too simple

\b Unitarity does not constrain lepton or photon-induced cross sections

\b We still cannot calculate $\P\P$ exchange -- even after more than 45 years

\b There are severe mathematical problems with DGLAP at small $x$

\b DGLAP cannot be used below $Q^2=5$ GeV$^2$

\b Regge fits to $F_2(x,Q^2)$ are the simplest --- and probably the most
correct

\b Elastic scattering at the LHC at large $t$ may be surprisingly large

\b If the CDF Tevatron cross section is correct, something
dramatic must happen --- independently of any theory!

\bibliographystyle{plainer}

\end{document}